\documentclass[twocolumn,english,prl,amsmath,amssymb,showpacs]{revtex4-1}
\usepackage[T1]{fontenc}
\usepackage[latin9]{inputenc}
\usepackage{amsmath}
\usepackage{amssymb}
\usepackage{esint}
\usepackage{pdfpages}

\makeatletter

\@ifundefined{textcolor}{}
{%
 \definecolor{BLACK}{gray}{0}
 \definecolor{WHITE}{gray}{1}
 \definecolor{RED}{rgb}{1,0,0}
 \definecolor{GREEN}{rgb}{0,1,0}
 \definecolor{BLUE}{rgb}{0,0,1}
 \definecolor{CYAN}{cmyk}{1,0,0,0}
 \definecolor{MAGENTA}{cmyk}{0,1,0,0}
 \definecolor{YELLOW}{cmyk}{0,0,1,0}
}

\usepackage{babel}
\usepackage{graphicx}

\makeatother

\usepackage{babel}
\begin{document}

\title{Spin torque and Nernst effects in Dzyaloshinskii-Moriya ferromagnets}

\author{Alexey A. Kovalev}

\affiliation{Department of Physics and Astronomy and Nebraska Center for Materials
and Nanoscience, University of Nebraska, Lincoln, Nebraska 68588,
USA}

\author{Vladimir Zyuzin}

\affiliation{Department of Physics and Astronomy and Nebraska Center for Materials
and Nanoscience, University of Nebraska, Lincoln, Nebraska 68588,
USA}

\date{\today}
\begin{abstract}
We predict that a temperature gradient can induce a magnon-mediated intrinsic torque in systems with non-trivial magnon Berry curvature. 
With the help of a microscopic linear response theory of nonequilibrium
magnon-mediated torques and spin currents we identify the interband and intraband components that manifest in
ferromagnets with Dzyaloshinskii\textendash Moriya interactions and magnetic textures. To illustrate and assess the importance of such effects, we 
apply the linear response theory to the magnon-mediated spin Nernst and torque responses
in a kagome lattice ferromagnet.
\end{abstract}


\maketitle
Studies of the spin degree of freedom in spintronics \cite{Zutic:RoMP2004}
naturally extend to include the interplay between the energy and spin
flows in the field of spincaloritronics \cite{Bauer.Saitoh.ea:NM2012,Goennenwein:mar2012}.
Improved efficiency in interconversion between energy and spin \cite{Weiler:Prl2013}
could result in important applications, e.g., for energy harvesing,
cooling, and magnetization control \cite{Hatami.Bauer.ea:PRL2007,Bauer.Bretzel.ea:2010,Kovalev:PRB2009,Cahaya.Tretiakov.ea:APL2014,Kovalev:SSC2010,Kovalev.Guengoerdue:EEL2015}.
Magnetic insulators such as yttrium iron garnet (YIG) or Lu$_{2}$V$_{2}$O$_{7}$
offer a perfect playground for spincaloritronics where due to the
absence of electron continuum the dissipation can be lowered as only the spin and energy matter \cite{Kajiwara.Harii.ea:2010,Onose.Ideue.ea:S2010,Uchida.Xiao.ea:NM2010}. It has already been demonstrated in recent experiments that energy currents can be used for magnetization control \cite{Torrejon.Malinowski.ea:PRL2012,Jiang:PRL2013}. This opens new possibilities for applications of magnon-mediated torques in racetrack memories \cite{Parkin.Hayashi.ea:S2008,Brataas.Kent.ea:NM2012}, and even in quantum information manipulations \cite{Kim.Tewari.ea:PRB2015}.

As we show in this study, the magnon-mediated torque is closely related to the magnon-mediated thermal Hall effect. The latter has been observed in Lu$_{2}$V$_{2}$O$_{7}$
\cite{Onose.Ideue.ea:S2010} and explained by the Berry curvature
of magnon bands \cite{Matsumoto.Murakami:PRL2011,Matsumoto.Shindou.ea:PRB2014,Katsura.Nagaosa.ea:PRL2010}
where the physics is reminiscent of the anomalous Hall effect \cite{Sundaram:PRB1999}.
The possibility of the magnon edge currents and tunable topology of
the magnon bands has also been discussed in the context of magnetic
insulators \cite{Matsumoto.Murakami:PRL2011,Zhang.Ren.ea:PRB2013,Mook.Henk.ea:PRB2014,Mook.Henk.ea:PRB2014a}.
In a recent experiment, the magnon-mediated thermal Hall effect showed the
sign reversal with changes in temperature or magnetic field in the
kagome magnet Cu(1-3, bdc) \cite{Hirschberger.Chisnell.ea:PRL2015}.
Since magnons also carry spin it would be natural to also study how
spin currents can be generated from temperature gradients, i.e., the
spin Nernst effect, in materials with nontrivial topology of magnon
bands. However, both the magnon-mediated torque and the spin Nernst effect have not been addressed in systems with non-trivial magnon Berry curvature. Such calculations inevitably require generalizations of linear response methods developed in sixties and seventies \cite{Luttinger:PR1964,Smrcka.Streda:JPC1977} to bosonic systems and consideration of the spin current analog of the energy magnetization contribution \cite{Qin.Niu.ea:PRL2011}.
\begin{figure} \centerline{\includegraphics[clip,width=1\columnwidth]{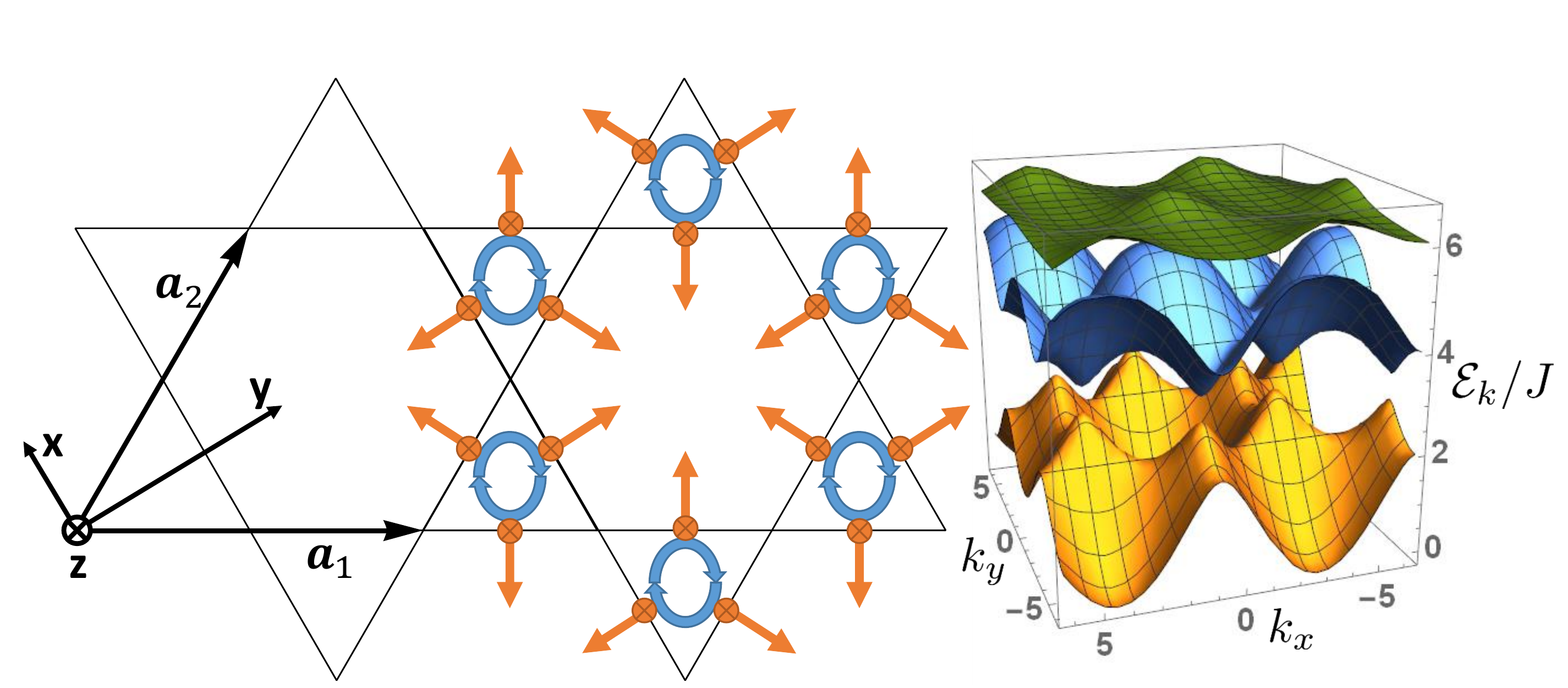}}
\protect\caption{(Color online) Left: Two-dimensional kagome lattice with lattice vectors $\boldsymbol{a}_1$ and $\boldsymbol{a}_2$ where atoms are placed in the corners of triangles. DMI are shown by vectors perpendicular to the bonds. The DMI of strength $D_1$ point into the page while the Rashba-like DMI of strength $D_2$ lie in the page. Right: The three magnon bands are plotted for the case of broken mirror symmetry with respect to the kagome plane due to the Rashba-like DMI. The direction of the spin density is given by $\mathbf{m}=\hat{x}\sin (\pi/6)+\hat{z} \cos (\pi/6)$. }
\label{fig:Fig1}  
\end{figure}

In this Rapid Communication, we predict that a temperature gradient can induce a magnon-mediated intrinsic torque in systems with non-trivial magnon Berry curvature.  To this end, we formulate a microscopic linear response theory
to temperature gradients for ferromagnets with multiple magnon bands.
We follow the Luttinger approach of the gravitational scalar potential
\cite{Luttinger:PR1964,Tatara:PRB2015}.
Our theory is capable of capturing the nontrivial topology of magnon
bands resulting from the Dzyaloshinskii\textendash Moriya
interactions (DMI) \cite{Moriya:PR1960,Dzyaloshinsky:JoPaCoS1958}. An additional vector potential corresponding to the magnetic texture can be readily introduced in our approach via minimal coupling.
 We note that the predicted magnon-mediated torques are bosonic analogs of the spin-orbit torques \cite{Bernevig.Vafek:PRB2005,Manchon.Zhang:PRB2008,Matos-Abiague.Rodriguez-Suarez:PRB2009,Chernyshov.Overby.ea:NP2009,Garate.MacDonald:PRB2009,Endo.Matsukura.ea:APL2010,Fang.Kurebayashi.ea:NN2011,Liu.Pai.ea:S2012,Kurebayashi.Sinova.ea:NN2014,Garello.Miron.ea:NN2013}.
We find that torques due to Dzyaloshinskii\textendash Moriya interactions (DM torques) can only
arise in systems lacking the center of inversion. This is in contrast to the the magnon-mediated spin Nernst effect. Finally, we express the intrinsic contribution to the DM torque via the mixed Berry curvature calculated with respect to the variation
of the magnetization and momentum \cite{Sundaram:PRB1999}. 
We apply our linear response theory to the magnon-mediated spin Nernst and
torque responses in a kagome lattice ferromagnet. We note that the latter can be detected by studying the magnetization dynamics while the former can be detected by the inverse spin Hall effect.

\textit{Microscopic theory.---} We consider a noninteracting boson
Hamiltonian describing the magnon fields:
\begin{equation}
\mathcal{H}=\int d\mathbf{r}\Psi^{\dagger}(\mathbf{r})H\Psi(\mathbf{r}),\label{eq:Ham}
\end{equation}
where $H$ is a Hermitian matrix of the size $N\times N$ and $\Psi^{\dagger}(\mathbf{r})=[a_{1}^{\dagger}(\mathbf{r}),\ldots,a_{N}^{\dagger}(\mathbf{r})]$)
describes $N$ bosonic fields corresponding to the number of modes
within a unit cell (or the number of spin-wave bands). Hamiltonian in Eq.~(\ref{eq:Ham}) can also account for smooth magnetic textures via minimal coupling to the texture-induced vector potential $\boldsymbol{\cal{A}}$ via additional term $(\boldsymbol{\cal{A}}_\alpha\cdot\mathbf{m})j^s_\alpha$ where $j^s_\alpha$ is the magnon spin current \cite{Kovalev:EPL2012,Tatara:PRB2015}. The Fourier
transformed Hamiltonian is:
\begin{equation}
\mathcal{H}=\sum_{\mathbf{k}}a_{k}^{\dagger}H(\mathbf{k})a_{k},\label{eq:Ham1}
\end{equation}
where $a_{k}^{\dagger}$ is the Fourier transformed vector of creation
operators. Hamiltonian in Eq.~(\ref{eq:Ham1}) can be diagonalized
by a unitary matrix $T_{k}$, i.e. $\mathcal{E}_{k}=T_{k}^{\dagger}H(\mathbf{k})T_{k}$
and $T_{k}^{\dagger}T_{k}=1_{N\times N}$ where $\mathcal{E}_{k}$
is the diagonal matrix of band energies, and $1_{N\times N}$ is the
$N\times N$ unit matrix. As it was shown by Luttinger \cite{Luttinger:PR1964},
the effect of the temperature gradient can be replicated by introducing
a perturbation to Hamiltonian in Eq.~(\ref{eq:Ham}):
\begin{equation}
\mathcal{H}^{'}=\dfrac{1}{2}\int d\mathbf{r}\Psi^{\dagger}(\mathbf{r})\left(H\chi+\chi H\right)\Psi(\mathbf{r}),\label{eq:Luttinger}
\end{equation}
where the nonequilibrium magnon-mediated field can be treated as a linear
response to the perturbation in Eq.~(\ref{eq:Luttinger}) and $\partial_{i}\chi=\partial_{i}T/T$. 

The nonequilibrium magnon-mediated field can be calculated by invoking arguments
similar to those for the spin-orbit torque \cite{Garate.MacDonald:PRB2009,Geranton.Freimuth.ea:PRB2015,Qaiumzadeh.Duine.ea:PRB2015}:
\begin{equation}
\mathbf{h}_{\text{tot}}=\mathbf{h}+\mathbf{h}^{'}=-\bigl\langle\partial_{\mathbf{m}}\mathcal{H}\bigr\rangle_{ne}-\bigl\langle\partial_{\mathbf{m}}\mathcal{H}^{'}\bigr\rangle_{eq},\label{eq:NEfields}
\end{equation}
where the averaging is done either over the equilibrium or nonequilibrium
state induced by the temperature gradient, and $\mathbf{m}$ is a
unit vector in the direction of the spin density $s$. The magnon-mediated torque
can be expressed as $\boldsymbol{\mathcal{T}}=\mathbf{m}\times\mathbf{h}_{\text{tot}}$ leading to modification of the Landau-Lifshitz-Gilbert equation, i.e., $s (1+ \alpha \mathbf{m} \times) \dot {\mathbf{m}} = \mathbf{m} \times\mathbf{H}_\text{eff}+\boldsymbol{\mathcal{T}}$ where $\mathbf{H}_\text{eff}$ is the effective magnetic field.
We are also concerned with the magnon current carrying spin which
has two components: 
\begin{equation}
\mathbf{J}_{\text{tot}}=\bigl\langle\mathbf{J}\bigr\rangle_{ne}+\bigl\langle\mathbf{J}^{'}\bigr\rangle_{eq},\label{eq:NEcurrent}
\end{equation}
where the first component, $\mathbf{J}=\int d\mathbf{r}\Psi^{\dagger}(\mathbf{r})\mathbf{v}\Psi(\mathbf{r})$,
does not depend on the temperature gradient and the second component,
$\mathbf{J}^{'}=(1/2)\int d\mathbf{r}\Psi^{\dagger}(\mathbf{r})\left(\mathbf{v}\chi+\chi\mathbf{v}\right)\Psi(\mathbf{r})$,
is proportional to the temperature gradient. The latter contribution is related to the spin current analog of the energy magnetization \cite{Qin.Niu.ea:PRL2011}. Here the velocity operator
is given by $\mathbf{v}=(1/i\hbar)[\mathbf{r},H]$. The magnon current
density $\mathbf{j}$ is introduced in a standard way from the continuity
equation $\dot{\rho}+\boldsymbol{\nabla}\cdot\mathbf{j}(\mathbf{r})=0$
where $\rho$ is the density of magnons. In our discussion, we employ
the expression for the energy current density, $\mathbf{j}^{Q}(\mathbf{r})=(1/2)\Psi^{\dagger}(\mathbf{r})(\mathbf{v}H+H\mathbf{v})\Psi(\mathbf{r})$,
and the macroscopic energy current $\mathbf{J}^{Q}=\int d\mathbf{r}\mathbf{j}^{Q}(\mathbf{r})$
corresponding to the continuity equation $\dot{\rho}_{E}+\boldsymbol{\nabla}\cdot\mathbf{j}^{Q}(\mathbf{r})=0$
with $\rho_{E}$ being the energy density. Note that we omitted the
component of $\mathbf{j}^{Q}$ proportional to $\partial_{i}\chi$
as it is irrelevant to our discussion. Within the linear response
theory, the response  of an operator X to temperature gradient becomes:
\begin{equation}
\bigl\langle X_{i}\bigr\rangle_{ne}=\lim_{\Omega\rightarrow0}\left\{ [\Pi_{ij}^{R}(\Omega)-\Pi_{ij}^{R}(0)]/i\Omega\right\} \partial_{j}\chi,\label{eq:field}
\end{equation}
where $\mathbf{X}$ is either spin current $-\hbar\mathbf{J}$ or
nonequilibrium field $\mathbf{h}=-\partial_{\mathbf{m}}H$, $\Pi_{ij}^{R}(\Omega)=\Pi_{ij}(\Omega+i0)$
is the retarded correlation function related to the following correlator in Matsubara formalism, $\Pi_{ij}(i\Omega)=-\int_{0}^{\beta}d\tau e^{i\Omega\tau}\bigl\langle T_{\tau}X_{i}J_{j}^{Q}\bigr\rangle$.
Note that the energy current originates from the expression $\mathcal{\dot{H}}^{'}=(i/\hbar)[\mathcal{H},\mathcal{H}^{'}]=\mathbf{J}^{Q}\boldsymbol{\partial}\chi$. 

We calculate the correlator in Eq.~(\ref{eq:field}) by considering
the simplest bubble diagram for $\Pi_{ij}$ and performing the analytic
continuation. We express the result through a response tensor $t_{ij}=t_{ij}^{\text{I}}+t_{ij}^{\text{II}}$
such that $X_{i}=-t_{ij}\partial_{j}\chi$ \cite{Note}: 
\begin{equation}
\begin{array}{c}
t_{ij}^{\text{I}}=\dfrac{1}{\hbar}{\displaystyle \int}\dfrac{d\omega}{2\pi}g(\omega)\dfrac{d}{d\omega}\mbox{Re}\mbox{Tr}\bigl\langle X_{i}G^{R}\mathcal{J}_{j}G^{A}-X_{i}G^{R}\mathcal{J}_{j}G^{R}\bigr\rangle,\\
t_{ij}^{\text{II}}=\dfrac{1}{\hbar}{\displaystyle \int}\dfrac{d\omega}{2\pi}g(\omega)\mbox{Re}\mbox{Tr}\bigl\langle X_{i}G^{R}\mathcal{J}_{j}\dfrac{dG^{R}}{d\omega}-X_{i}\dfrac{dG^{R}}{d\omega}\mathcal{J}_{j}G^{R}\bigr\rangle,
\end{array}\label{eq:field1}
\end{equation}
where $g(\omega)$ is the Bose distribution function $g(\omega)=1/[\exp(\hbar\omega/k_{B}T)-1]$,
$G^{R}=\hbar(\hbar\omega-H+i\Gamma)^{-1}$, $G^{A}=\hbar(\hbar\omega-H-i\Gamma)^{-1}$,
and $\boldsymbol{\mathcal{J}}=(\mathbf{v}H+H\mathbf{v})/2$. For practical
purposes, we Fourier transform Eq.~(\ref{eq:field1}) which leads to
additional momentum integration and momentum transformed terms, i.e.
$G^{R}(\mathbf{k})=\hbar(\hbar\omega-H(\mathbf{k})+i\Gamma)^{-1}$,
$\mathbf{h}_{k}=-\partial_{\mathbf{m}}H(\mathbf{k})$, $\mathbf{v}_{k}=\partial_{\hbar\mathbf{k}}H(\mathbf{k})$,
and $\boldsymbol{\mathcal{J}}{}_{k}=(\mathbf{v}_{k}H(\mathbf{k})+H(\mathbf{k})\mathbf{v}_{k})/2$.
The approximation we are using can be improved by performing the disorder
averaging which is indicated by brackets in Eq.~(\ref{eq:field1})
. In addition, interactions with phonons can be also taken into account
and can result in additional dissipative corrections to the torque.
Throughout this paper, we adopt a simple phenomenological treatment
by relating the quasiparticle broadening to the Gilbert damping, i.e.
$\Gamma=\alpha\hbar\omega$. 

\textit{Berry curvature formulation.---} It is very insightful to
carry out the frequency integrations in Eq.~(\ref{eq:field1}), keeping
only the two leading orders in $\Gamma$ and combining the linear
response result with the nonequilibrium contribution $\mathbf{h}^{'}$
in Eq.~(\ref{eq:NEfields}) or $\mathbf{J}^{'}$ in Eq.~(\ref{eq:NEcurrent}).
To carry the integrations in Eq.~(\ref{eq:field1}) we use the diagonal
basis defined by rotation matrices $T_{k}$, and transform the contributions
$\mathbf{h}^{'}$ and $\mathbf{J}^{'}$ to an integral over energies
following the approach of Smrcka and Streda \cite{Smrcka.Streda:JPC1977,Matsumoto.Shindou.ea:PRB2014}.
Using the covariant derivative we calculate the rotated velocity,
$T_{k}^{\dagger}\hbar\mathbf{v}_{k}T_{k}=\partial_{\mathbf{k}}\mathcal{E}_{k}-i\boldsymbol{\mathcal{A}}_{k}\mathcal{E}_{k}+i\mathcal{E}_{k}\boldsymbol{\mathcal{A}}_{k}$,
and nonequilibrium field, $T_{k}^{\dagger}\mathbf{h}_{k}T_{k}=\partial_{\mathbf{m}}\mathcal{E}_{k}-i\boldsymbol{\mathcal{A}}_{m}\mathcal{E}_{k}+i\mathcal{E}_{k}\boldsymbol{\mathcal{A}}_{m}$,
where $\boldsymbol{\mathcal{A}}_{k}=iT_{k}^{\dagger}\partial_{\mathbf{k}}T_{k}$
and $\boldsymbol{\mathcal{A}}_{m}=iT_{k}^{\dagger}\partial_{\mathbf{m}}T_{k}$.
Substituting these in Eq.~(\ref{eq:field1}) we identify the intraband
and interband contributions to the response tensor \cite{Note}:
\begin{equation}
\begin{array}{c}
t_{ij}^{\text{intra}}=\dfrac{1}{V}{\displaystyle \sum_{\mathbf{k}}{\displaystyle \sum_{n=1}^{N}}\dfrac{1}{2\Gamma_{k}}(\partial_{x_{i}}\varepsilon_{nk})(\partial_{k_{j}}\varepsilon_{nk})\varepsilon_{nk}g'(\varepsilon_{nk})},\\
t_{ij}^{\text{inter}}=\dfrac{k_{B}T}{V}{\displaystyle \sum_{\mathbf{k}}}{\displaystyle \sum_{n=1}^{N}}c_{1}[g(\varepsilon_{nk})]\Omega_{x_{i}k_{j}}^{n}(\mathbf{k}),
\end{array}\label{eq:field2}
\end{equation}
where $x_{i}$ is either $m_{i}$ or $k_{i}$, $\varepsilon_{nk}=[\mathcal{E}_{k}]_{nn}$,
$\Gamma_{nk}=\alpha\varepsilon_{nk}$, $g'(\varepsilon_{nk})=(2k_{B}T)^{-1}\{1-\cosh(\varepsilon_{nk}/k_{B}T)\}^{-1}$,
$c_{1}[x]=\int_{0}^{x}dt\ln[(1+t)/t]=(1+x)\ln[1+x]-x\ln x$, $V$
is volume, and we introduced the Berry curvature of $n$-th band:
\begin{equation}
\Omega_{x_{i}k_{j}}^{n}(\mathbf{k})=i[(\partial_{x_{i}}T_{k}^{\dagger})(\partial_{k_{j}}T_{k})-(\partial_{k_{j}}T_{k}^{\dagger})(\partial_{x_{i}}T_{k})]_{nn}.\label{eq:Berry}
\end{equation}
Such Berry curvatures naturally appear in discussions of semiclassical equations of  motion for Hamiltonians with slowly varying parameters \cite{Sundaram:PRB1999}. 
Derivation of Eq.~(\ref{eq:field2}) (see supplemental material \cite{Note}) should also hold for fermion systems given that $c_{1}(\varepsilon_{nk}) = - \int_{\varepsilon_{nk}}^{\infty} \eta\frac{d n_F(\eta)}{d\eta} d\eta$ where the Fermi-Dirac distribution $n_F$ replaces $g$ \cite{Freimuth.Bluegel.ea:apa2016}. 
By applying the time reversal transformation, i.e. $\mathbf{k}\rightarrow-\mathbf{k}$,
$\mathbf{m}\rightarrow-\mathbf{m}$, $\Omega_{x_{i}k_{j}}^{n}\rightarrow-\Omega_{-x_{i}-k_{j}}^{n}$,
to Eqs. (\ref{eq:field2}) we recover the transformation properties
of $t_{ij}^{\text{intra}}$ and $t_{ij}^{\text{inter}}$ under the
magnetization reversal. In particular, it is clear that $t_{ij}^{\text{intra}}$
is even under the magnetization reversal and is divergent as $\Gamma\rightarrow0$.
On the other hand, $t_{ij}^{\text{inter}}$ is odd under the magnetization
reversal and corresponds to the intrinsic contribution independent
of $\Gamma$. In terms of spin torques, the former corresponds to
the field-like torque and the latter to the anti-damping (or dissipative)
intrinsic torque.

\textit{Model.---} We apply our theory to the magnon current and torque
response of a kagome lattice ferromagnet with DMI (see Fig.~\ref{fig:Fig1}).
The exchange and DMI terms in the Hamiltonian are given by \cite{Moriya:PR1960,Dzyaloshinsky:JoPaCoS1958}:
\begin{equation}
{\cal H}=-\frac{1}{2}J\sum_{i\neq j}\mathbf{S}_{i}\cdot\mathbf{S}_{j}+\frac{1}{2}\sum_{i\neq j}\mathbf{D}_{ij}\cdot(\mathbf{S}_{i}\times\mathbf{S}_{j}),\label{eq:DMI}
\end{equation}
where $J$ corresponds to the nearest neighbor interaction, $\mathbf{D}_{ij}$
is the DMI vector between sites $i$ and \textbf{$j$} ($\mathbf{D}_{ij}=\mathbf{-}\mathbf{D}_{ji}$).
We take the DMI vector to be $\mathbf{D}_{ij}=D_{1}\hat{z}$ for the
ordering of sites shown by the arrow inside triangles in Fig.~\ref{fig:Fig1}.
Such configuration corresponds to systems with the center of inversion.
In some cases, we also add a Rashba-like inplane contribution, $\mathbf{D}_{ij}=D_{2}(\hat{z}\times\overrightarrow{ij})$,
that breaks the mirror symmetry with respect to the kagome plane where
$\overrightarrow{ij}$ is a unit vector connecting sites $i$ and
$j$ ($\mathbf{D}_{ij}$ is shown by arrows in Fig.~\ref{fig:Fig1}).
We also add the Zeeman term due to an external magnetic field that
fixes the direction of the magnetization direction along the field.
After applying the Holstein-Primakoff transformation, we arrive at
a noninteracting Hamiltonian compatible with Eq.~(\ref{eq:Ham}).
A typical magnon spectrum is shown in Fig.~\ref{fig:Fig1} where the
lower, middle, and upper bands have the Chern numbers $1$, $0$,
and $-1$, respectively. 

\begin{figure} \centerline{\includegraphics[clip,width=1\columnwidth]{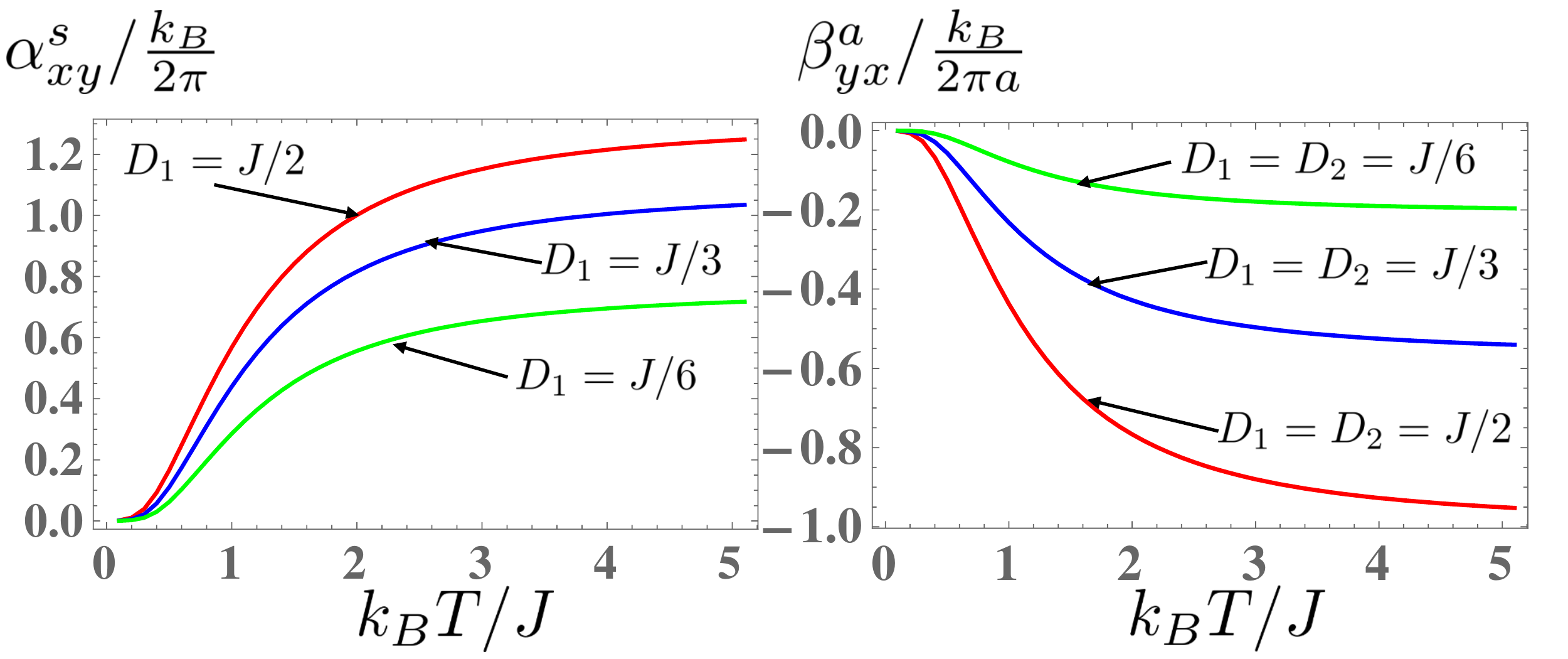}}
\protect\caption{(Color online) Left: The spin Nernst conductivity $\alpha_{ij}^{s}$ versus temperature $T$ for DMI $D_2=0$ and $D_1=J/2$, $J/3$, and $J/6$. Right: The thermal torkance $\beta_{yx}^a$ corresponding to the anti-damping part of the torque versus temperature $T$ for DMI   $D_1=D_2=J/2$, $J/3$, and $J/6$. Note that the temperature range is not limited by the Curie temperature in order to show the assymtotic behavior. In both figures the direction of the spin density is given by $\mathbf{m}=\hat{z}$. }
\label{fig:Fig2}  
\end{figure}

\textit{Spin Nernst effect.---} The thermal Hall effect manifests
itself in the transverse temperature gradient \cite{Onose.Ideue.ea:S2010,Matsumoto.Shindou.ea:PRB2014,Mook.Henk.ea:PRB2014a}.
Here we calculate the transverse spin current which can be detected,
e.g., via the inverse spin Hall effect in a Pt contact attached to
the sample \cite{Saitoh:APL2006}. The spin Nernst conductivity $\alpha_{ij}^{s}$
relates the temperature gradient to the spin current density, i.e.
$j_{i}^{s}=-\hbar j_{i}=-\alpha_{ij}^{s}\partial_{j}T$ where each
magnon carries the angular momentum $-\hbar$. From Eq.~(\ref{eq:field2})
we obtain $\alpha_{ij}^{s}=t_{ij}/T$ with only the interband part
contributing to $\alpha_{ij}^{s}$. For a model calculation, we consider
Eq.~(\ref{eq:DMI}). The spin Nernst effect can take place in systems
with the center of inversion, thus the Rashba-like DMI described by
parameter $D_{2}$ can be zero. By integrating the Berry curvature
over the Brillouin zone, we arrive at the result in Fig.~\ref{fig:Fig2}
where $\alpha_{ij}^{s}$ is dominated by the lowest band in Fig.~\ref{fig:Fig1}
with the positive Chern number. For a three-dimensional system containing
weakly interacting kagome layers, we can write $\alpha_{ij}^{\text{3D}}=\alpha_{ij}^{s}/c$
where $c\propto a$ is the interlayer distance and $a$ is the lattice
constant. Given results in Fig.~\ref{fig:Fig2}, it seems to be possible
to generate a transverse spin current of the order of $10^{-10}\mbox{J}/\mbox{m}^{2}$
from a temperature gradient of $20\,\mbox{K}/\mbox{mm}$ \cite{Jiang:PRL2013}
in three dimensional systems. Spin currents of such magnitude are
typical for spin pumping experiments \cite{Weiler:Prl2013}.

\begin{figure} \centerline{\includegraphics[clip,width=1\columnwidth]{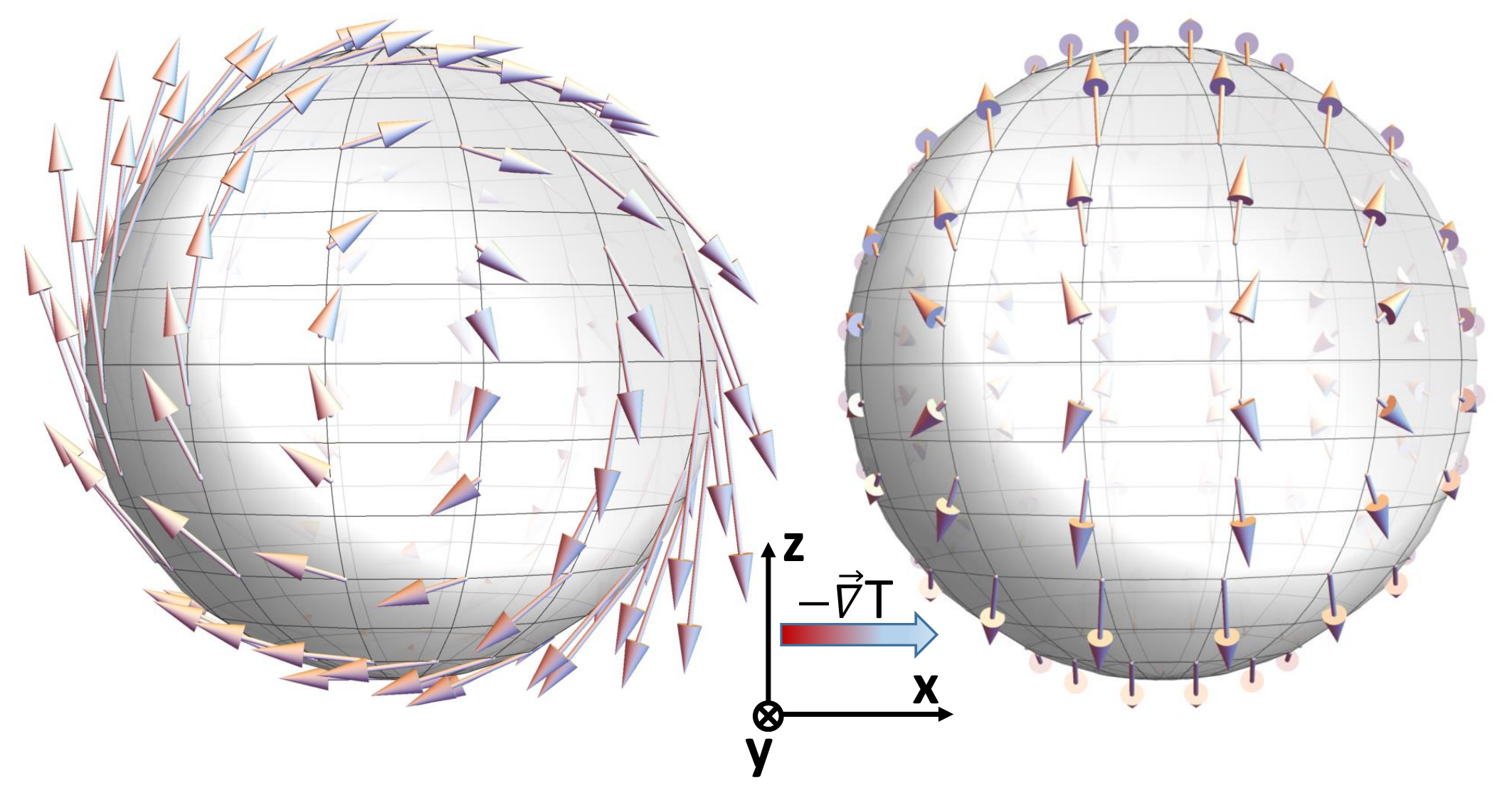}}
\protect\caption{(Color online) The nonequilibrium magnon-mediated torque $\boldsymbol{\mathcal{T}}$ is plotted on a unit sphere representing the direction of uniform spin density $\mathbf{m}$. The temperature is $T=2J$ and the gradient is applied along the $-\hat{x}$ direction. The field-like torque component $\boldsymbol{\mathcal{T}}^f$ that is odd in the magnetization is plotted on the left and the anti-damping component $\boldsymbol{\mathcal{T}}^a$ that is even in the magnetization is plotted on the right. The field-like component is rescaled by the Gilbert damping to match in scale the anti-damping component, i.e., $\boldsymbol{\mathcal{T}}^f \rightarrow \alpha\boldsymbol{\mathcal{T}}^f$.}
\label{fig:Fig3}  
\end{figure}

\textit{Nonequilibrium torques.---} To present our results we introduce
the thermal torkance $\beta_{ij}$ that relates the magnetization
torque to the temperature gradient, i.e. $\mathcal{T}_{i}=-\beta_{ij}\partial_{j}T$
or $\beta_{ij}=m_{l}\varepsilon_{lki}t_{kj}/T$ in terms of Eq.~(\ref{eq:field2})
where $\varepsilon_{lki}$ is the antisymmetric tensor. We further
separate the torkance $\beta_{ij}$ into the field-like part $\beta_{ij}^{f}$
that is odd in the magnetization and the anti-damping part $\beta_{ij}^{a}$
that is even in the magnetization. 

To uncover the effect of Berry curvature, we apply our theory to the
model in Eq.~(\ref{eq:DMI}). Within our theory the anti-damping component
of the torque entirely comes from the Berry curvature contribution
in Eq.~(\ref{eq:field2}). The largest component of $\beta_{ij}^{a}$
corresponding to the temperature gradient along the $x-$axis, the
torque along the $y-$axis, and the spin density along the $z-$axis
is plotted in Fig.~\ref{fig:Fig2}. The temperature dependence of
$\beta_{ij}^{a}$ resembles the temperature dependence of the spin
Nernst conductivity where we observe larger effect at higher temperatures.
For a three-dimensional system containing weakly interacting kagome
layers, we obtain $\beta_{ij}^{\text{3D}}=\beta_{ij}/c$ where $c$
is the interlayer distance. In Fig.~\ref{fig:Fig3}, we plot the nonequilibrium
magnon-mediated torque separated into the field-like and anti-damping parts,
$\boldsymbol{\mathcal{T}}=\boldsymbol{\mathcal{T}}^{f}+\boldsymbol{\mathcal{T}}^{a}$,
on a unit sphere representing the spin density vector $\mathbf{m}$.
The torque in Fig.~\ref{fig:Fig3} can be obtained from phenomenological
expressions obtained for films with structural asymmetry along the
$z-$axis \cite{Manchon.Ndiaye.ea:PRB2014,Kovalev.Guengoerdue:EEL2015},
$\boldsymbol{\mathcal{T}}_{i}^{f}\propto(\mathbf{m}\times\mathbf{D}_{i})\partial_{i}T$
and $\boldsymbol{\mathcal{T}}_{i}^{a}\propto\mathbf{m}\times(\mathbf{m}\times\mathbf{D}_{i})\partial_{i}T$,
by a deformation not involving the change in topology where $\mathbf{D}_{i}=\mathbf{e}_{z}\times\mathbf{e}_{i}$
and $i$ is either $x$ or $y$.

A ballpark estimate of the strength of the nonequilibrium magnon-mediated
torque can be done by considering only the lowest band in the quadratic
approximation, i.e., we have $H(\mathbf{k})=\hbar A[k_{\alpha}+\mathbf{m}\cdot(\mathbf{D}_{\alpha}/A-\boldsymbol{\cal{A}}_\alpha)]^{2}/s$
where $A$ is the exchange stiffness, $\boldsymbol{\cal{A}}_\alpha$ is the texture-induced vector potential, $s$ is the spin density, and
a tensor $D_{\alpha\beta}=\mathbf{D}_{\alpha}\cdot\mathbf{e}_{\beta}$
describes DMI. After substituting this spectrum in the first Eq.~(\ref{eq:field2})
we obtain the longitudinal spin current $\mathbf{j}^{s}=-\hbar\mathbf{j}=k_{B}\boldsymbol{\partial}T[\sqrt{\pi}\zeta(3/2)]/(8\pi^{2}\lambda\alpha)$
where $\zeta$ is the Riemann zeta function and $\lambda=\sqrt{\hbar A/sk_{B}T}$
is the thermal magnon wavelength. The same Eq.~(\ref{eq:field2})
results in the expression for the nonequilibrium field-like torque
density: 
\begin{equation}
\boldsymbol{\mathcal{T}}^{f}=[\mathbf{m}\times(\mathbf{D}_{\alpha}/A-\boldsymbol{\cal{A}}_\alpha)]j_{\alpha}^{s},\label{eq:torque}
\end{equation}
 which agrees with the earlier results obtained for a single-band ferromagnet
\cite{Kovalev:PRB2014,Manchon.Ndiaye.ea:PRB2014,Linder:PRB2014,Kovalev.Guengoerdue:EEL2015,Kim.Tserkovnyak:PRB2015}. Here the torque is generated within the whole volume. This is contrary
to the conventional spin-transfer torque which is generated only close to the interface \cite{Kovalev.Brataas.ea:PRB2002}. The typical
charge current density $j^{e}=10^{10}\mbox{A}/\mbox{\ensuremath{m^{2}}}$
sufficient for the spin-transfer torque switching should be compared
to $2ejdD/A\approx10^{9}\mbox{A}/\mbox{\ensuremath{m^{2}}}$ where
$e$ is the electron charge, $D$ is the strength of DMI and $d$
is the width of the magnet. For the estimate of the field-like torque, we assume that $d=A/D$,
$\partial_{i}T=20\,\mbox{K}/\mbox{mm}$, and $\alpha=10^{-4}$ \cite{Jiang:PRL2013}.

\textit{Conclusions.--- }We developed a linear response theory to
temperature gradients for magnetization
torques (DM torques). We identify the intrinsic part of the DM torque
and express it through the Berry curvature. We note that similar expressions
also arise for the magnon-mediated spin Nernst effect. According to our estimates,
the spin Nernst effect leads to substantial spin currents that could
be measured, e.g., by the inverse spin Hall techniques \cite{Saitoh:APL2006}
in such materials as pyrochlore crystals (e.g., Lu$_{2}$V$_{2}$O$_{7}$)
and the kagome ferromagnets \cite{Hirschberger.Chisnell.ea:PRL2015,Boldrin.Fak.ea:PRB2015} {[}e.g., Cu(1-3, bdc){]}. In particular,
a voltage should arise in the neighboring heavy metal due to the inverse
spin Hall effect in full analogy to measurements of the spin Seebeck
effect and spin pumping \cite{Weiler:Prl2013}. We also find that
the DM torques should influence the magnetization dynamics in ferromagnets
with DMI; however, larger temperature gradients (compared to $20\,\mbox{K}/\mbox{mm}$ used in estimates \cite{Jiang:PRL2013}) are required, e.g., for magnetization
switching \cite{Pushp.Phung.ea:PNASU2015}. For the validity of the linear response approximation the temperature should not change much over the magnon mean free path. The DM torque can only
arise in materials with structural asymmetry or lacking the center
of inversion. Of relevance could be jarosites \cite{Elhajal.Canals.ea:PRB2002}
or ferromagnets and ferrimagnets containing buckled kagome layers
\cite{Pregelj.Zaharko.ea:PRB2012,Rousochatzakis.Richter.ea:PRB2015}.
Our theory can be readily generalized to antiferromagnets and ferrimagnets,
extending the range of materials suitable for observation of DM torques.
In particular, antiferromagnet does not have to have the center of
inversion in order to exhibit the DM torque provided each sublattice
individually lacks the center of inversion. 

We gratefully acknowledge stimulating discussions with Kirill Belashchenko and
Gen Tatara. This work was supported in part by the DOE Early Career
Award DE-SC0014189, and by the NSF under Grants Nos. Phy-1415600, PHY11-25915, and
DMR-1420645.

\bibliographystyle{apsrev}
\bibliography{MyBIB}


\section*{Supplementary material}

# Supplemental Material for
# Spin torque and Nernst effects in Dzyaloshinskii-Moriya ferromagnets

Alexey A. Kovalev and Vladimir Zyuzin
*Department of Physics and Astronomy and Nebraska Center for Materials and Nanoscience,*
*University of Nebraska, Lincoln, Nebraska 68588, USA*

### A. The Kubo formula for linear response

We study a response of an operator $X$ to the perturbation described by a Hamiltonian $\mathcal{H}'$. The total Hamiltonian of the system is:

$$\mathcal{H}_{\text{tot}} = \mathcal{H} + \mathcal{H}', \tag{1}$$

where $\mathcal{H} = \int d\mathbf{r} \Psi^\dagger(\mathbf{r}) H \Psi(\mathbf{r})$ and $\mathcal{H}' = \int d\mathbf{r} \Psi^\dagger(\mathbf{r}) H' \Psi(\mathbf{r})$. We assume that the Hamiltonian $H$ does not contain particle non-conserving (pairing) terms. The perturbing Hamiltonian could be $H' = \frac{1}{2}(\chi H + H\chi)$ which corresponds to the temperature gradient $\partial_i \chi = \partial_i T/T$, and $H' = e\phi$ which corresponds to the electric field $E_i = -\partial_i \phi$ with $e$ being the charge.

We assume that the operator $X$ can in general depend on the gradients. We are going to consider only the responses of the global operators $A \equiv \frac{1}{V} \int d\mathbf{r} \Psi^\dagger(\mathbf{r}) X \Psi(\mathbf{r})$ where we take $V = 1$ and reintroduce $V$ in the final expressions. Within the linear response, we only need to consider the two lowest orders with respect to gradients:

$$A = A^{(0)} + A^{(1)}. \tag{2}$$

Let us now write a general expression for the response of an operator $A$ to the perturbation $\mathcal{H}'$:

$$\langle A \rangle = \left\langle A^{(0)} \right\rangle_{ne} + \left\langle A^{(1)} \right\rangle_{eq}, \tag{3}$$

where $<..>_{ne}$ is the average over the non-equilibrium states due to the gradient terms, and $<..>_{eq}$ is the average over the equilibrium state. The first average can be found from the Kubo formula of the linear response [1]. It is given by the correlation function:

$$\left\langle A^{(0)}(\omega) \right\rangle_{ne} = \frac{i}{\hbar} \int_0^\infty \left\langle [\hat{A}(t), \mathcal{H}'(0)] \right\rangle e^{i(\omega+i0^+)t} dt \tag{4}$$

where $\hat{A}(t) = \sum_{mk} A_{mk} \Psi_m^\dagger \Psi_k e^{-i(E_k-E_m)t}$ and $\mathcal{H}'(t) = \sum_{mk} H'_{mk} \Psi_m^\dagger \Psi_k e^{-i(E_k-E_m)t}$ are the eigen-basis representations, with the field operators satisfying the relation $H\Psi_n(\mathbf{r}) = E_n \Psi_n(\mathbf{r})$. We use the relation $\dot{\mathcal{H}}' = (i/\hbar)[\mathcal{H}, \mathcal{H}'] = \mathbf{J}^Q \partial \chi$ when integrating by parts to obtain:

$$\left\langle A^{(0)}(\omega) \right\rangle_{ne} = \frac{i}{\hbar} \int_0^\infty \left\langle [\hat{A}(0), \frac{\partial \mathcal{H}'(-t)}{\partial t}] \right\rangle \frac{1}{i\omega - 0^+} dt - \frac{i}{\hbar} \int_0^\infty \left\langle [\hat{A}(0), \frac{\partial \mathcal{H}'(-t)}{\partial t}] \right\rangle \frac{e^{i(\omega+i0^+)t}}{i\omega - 0^+} dt \tag{5}$$

$$\equiv \frac{1}{i\omega - 0^+} \partial \chi \left[ \mathbf{K}_0(\omega) - \mathbf{K}_0(0) \right], \tag{6}$$

where we assumed that the perturbation vanishes as $t \to -\infty$. We also introduces a new correlator:

$$\mathbf{K}_0(\omega) = \frac{i}{\hbar} \int_0^\infty \left\langle [\hat{A}(t), \mathbf{J}^Q] \right\rangle e^{i(\omega+i0^+)t} dt. \tag{7}$$

To proceed further in simplifying the correlator, we use the following identities:

$$\langle \Psi_m^\dagger \Psi_n \Psi_p^\dagger \Psi_q \rangle - \langle \Psi_p^\dagger \Psi_q \Psi_m^\dagger \Psi_n \rangle = \delta_{mq} \delta_{np} n(E_m)[1 + n(E_n)] - \delta_{mq} \delta_{np} n(E_n)[1 + n(E_m)], \tag{8}$$

where $\langle ... \rangle$ stands for taking an average over the unperturbed state, $n(E) = 1/(e^{\beta E} - 1)$ is the Bose distribution function, and $\beta = 1/T$. After some transformations, we obtain:

$$\mathbf{K}_0(\omega) = \frac{1}{\hbar} \sum_{n,m} \frac{n(E_m) - n(E_n)}{E_n - E_m - \omega - i0^+} A_{mn}^{(0)} \mathbf{J}_{nm}^Q, \tag{9}$$

which is equivalent to the following expression:

$$\mathbf{K}_0(\omega) = \frac{1}{\hbar} \sum_{nm} \int_{-\infty}^{\infty} dE\, n(E) \left[ \frac{\delta(E - E_m)}{E_n - E - \omega - i0^+} - \frac{\delta(E - E_n)}{E - E_m - \omega - i0^+} \right] A_{mn}^{(0)} \mathbf{J}_{nm}^Q. \tag{10}$$

The static limit can be calculated by expanding the last expression in $\omega$. The zeroth order term in $\omega$ cancels with the $\mathbf{K}(0)$ contribution to the correlation function which results in the expression:

$$\mathbf{K}_0(\omega) - \mathbf{K}_0(0) \approx \frac{\omega}{\hbar} \sum_{nm} \int_{-\infty}^{\infty} dE\, n(E) \left[ \frac{\delta(E - E_m)}{(E - E_n + i0^+)^2} - \frac{\delta(E - E_n)}{(E - E_m - i0^+)^2} \right] A_{mn}^{(0)} \mathbf{J}_{nm}^Q. \tag{11}$$

This expression coressponds to the one derived by Crépieux and Bruno in [2]. Identities $\delta(E - E_n) = -(2\pi i)^{-1} \left( G_n^{\mathrm{R}}(E) - G_n^{\mathrm{A}}(E) \right)$ and $(E - E_n \pm i0^+)^{-2} = -\frac{\partial G_n^{\mathrm{R/A}}(E)}{\partial E}$ are of use in order to express the correlator in terms of Green's functions:

$$\left\langle A^{(0)} \right\rangle_{ne} = \frac{1}{2\pi\hbar} \sum_{nm} \int_{-\infty}^{+\infty} dE\, n(E) \left[ \left( \frac{\partial G_n^{\mathrm{R}}}{\partial E} \right) (G_m^{\mathrm{R}} - G_m^{\mathrm{A}}) - \left( \frac{\partial G_m^{\mathrm{A}}}{\partial E} \right) (G_n^{\mathrm{R}} - G_n^{\mathrm{A}}) \right] A_{mn}^{(0)} \mathbf{J}_{nm}^Q \boldsymbol{\partial}\chi. \tag{12}$$

This expression is consistent with the expression (31) in Smrcka and Streda in [3]. After some transformations, we recover expression corresponding to Eq. (7) from the main text:

$$\left\langle A^{(0)} \right\rangle_{ne} = \int_{-\infty}^{+\infty} \frac{dE}{2\pi\hbar} \left( n(E) \mathrm{Re} \frac{\partial}{\partial E} \left\{ [G_n^{\mathrm{R}} G_m^{\mathrm{A}} - G_n^{\mathrm{R}} G_m^{\mathrm{R}}] A_{mn}^{(0)} \mathbf{J}_{nm}^Q \right\} - n(E) \mathrm{Re} \left\{ \left[ \frac{\partial G_n^{\mathrm{R}}}{\partial E} G_m^{\mathrm{R}} - G_n^{\mathrm{R}} \frac{\partial G_m^{\mathrm{R}}}{\partial E} \right] A_{mn}^{(0)} \mathbf{J}_{nm}^Q \right\} \right) \boldsymbol{\partial}\chi. \tag{13}$$

Here, we assumed that the boundary terms vanish, and possible singularity at $E = 0$ can be regularized.

It is instructive to show that the same expressions are obtained when calculating the response correlation function in the Matsubara frequency. The correlator becomes:

$$\langle A(\omega_m) \rangle_{\mathrm{ne}} = \int_0^\beta d\tau e^{i\omega_m \tau} \left\langle T_\tau A^{(0)}(0) \mathcal{H}'(-\tau) \right\rangle, \tag{14}$$

where $\beta = 1/T$, $\omega = 2\pi T m$ is the boson Matsubara frequency with integer $m$, and $\tau = it$ is the imaginary time. We then perform integration by parts:

$$\frac{1}{i\omega_m} \int_0^\beta d\tau \frac{de^{i\omega_m \tau}}{d\tau} \left\langle T_\tau A^{(0)}(0) \mathcal{H}'(-\tau) \right\rangle = \frac{e^{i\omega_m \tau}}{i\omega_m} \left\langle T_\tau A^{(0)}(0) \mathcal{H}'(-\tau) \right\rangle |_0^\beta \tag{15}$$

$$- \int_0^\beta d\tau \frac{e^{i\omega_m \tau}}{i\omega_m} \left\langle T_\tau A^{(0)}(0) \frac{\partial \mathcal{H}'(-\tau)}{\partial \tau} \right\rangle. \tag{16}$$

We will use the relation $i\partial_\tau \mathcal{H}' = (i/\hbar)[\mathcal{H}, \mathcal{H}'] = \mathbf{J}^Q \boldsymbol{\partial}\chi$. Observing that $e^{i\omega_m \beta} = 1$, we obtain the correlator:

$$\frac{e^{i\omega_m \tau}}{i\omega_m} \left\langle T_\tau A^{(0)}(0) \mathcal{H}'(-\tau) \right\rangle |_0^\beta = \int_0^\beta d\tau \frac{1}{i\omega_m} \left\langle T_\tau A^{(0)}(0) \frac{\partial \mathcal{H}'(-\tau)}{\partial \tau} \right\rangle \equiv \frac{1}{i\omega_m} \mathbf{K}(0) \boldsymbol{\partial}\chi. \tag{17}$$

Therefore, for $\omega_m \to 0$ we can write

$$\langle A(\omega) \rangle_{ne} = \frac{1}{i\omega_m} \left( \mathbf{K}(0) - \mathbf{K}(\omega_m) \right) \boldsymbol{\partial}\chi = i \left[ \frac{\partial}{\partial \omega_m} \mathbf{K}(\omega_m) \boldsymbol{\partial}\chi \right]_{\omega_m \to 0}. \tag{18}$$

### B. Total current response to temperature gradient

Magnon current corresponding to $\rho(\mathbf{r}) = \Psi^\dagger(\mathbf{r})\Psi(\mathbf{r})$ density, is given

$$\mathbf{j} = \tilde{\Psi}^\dagger(\mathbf{r}) \mathbf{v} \tilde{\Psi}(\mathbf{r}), \tag{19}$$

where $\tilde{\Psi}^\dagger(\mathbf{r}) = \left(1 + \frac{\mathbf{r}\chi}{2}\right)\Psi(\mathbf{r})$, $\mathbf{v} = i[\mathcal{H}, \mathbf{r}]$ is the velocity and we defined $\mathbf{j}^{[0]} = \Psi^\dagger(\mathbf{r})\mathbf{v}\Psi(\mathbf{r})$ and $\mathbf{j}^{[1]} = \frac{1}{2}\Psi^\dagger(\mathbf{r})\left(\mathbf{v}r_\beta + r_\beta\mathbf{v}\right)\Psi(\mathbf{r})\nabla_\beta\chi$. We will be working with macroscopic currents, defined as $\mathbf{J} = \int d\mathbf{r}\mathbf{j}$. When calculating the linear response, we define

$$\langle J_\alpha \rangle = \left\langle J_\alpha^{[0]} \right\rangle_{\text{ne}} + \left\langle J_\alpha^{[1]} \right\rangle_{\text{eq}}, \tag{20}$$

where the first term is an average of $J_\alpha^{[0]}$ over the non-equilibrium state, which is just a Kubo linear response expression. We then write

$$J_\alpha^{[0]} = -\lim_{\omega \to 0}\frac{\partial}{\partial\omega}\Pi_{\alpha\beta}(\omega) \equiv S_{\alpha\beta}\nabla_\beta\chi, \tag{21}$$

where $\Pi_{\alpha\beta} = \int_0^{1/T} d\tau e^{i\omega\tau}\left\langle T_\tau J_\alpha^{[0]}(\tau)J_\beta^Q(0)\right\rangle$, where $\omega$ is a boson Matsubara frequency, $T$ is temperature, and $J_\beta^Q = \frac{1}{2}\int d\mathbf{r}\Psi^\dagger(\mathbf{r})[Hv_\beta + v_\beta H]\Psi(\mathbf{r})\nabla_\beta\chi$ is the current occuring when deriving the Kubo formula. The other term in the definition of the response $\langle J_\alpha\rangle$ has a straightforward definition

$$\left\langle J_\alpha^{[1]}\right\rangle_{\text{eq}} = \frac{1}{2V}\int d\mathbf{r}\left\langle \Psi^\dagger(\mathbf{r})\left(\mathbf{v}r_\beta + r_\beta\mathbf{v}\right)\Psi(\mathbf{r})\right\rangle\nabla_\beta\chi \equiv M_{\alpha\beta}\nabla_\beta\chi. \tag{22}$$

We then write

$$\langle J_\alpha\rangle = [S_{\alpha\beta} + M_{\alpha\beta}]\nabla_\beta\chi \equiv L_{\alpha\beta}\nabla_\beta\chi. \tag{23}$$

Let us deinfe a unitary matrix that diagonalizes the Hamiltonian. In Fourier space, we write

$$T_\mathbf{k}^\dagger H_\mathbf{k} T_\mathbf{k} = \mathcal{E}_\mathbf{k}, \tag{24}$$

where $T_\mathbf{k}^\dagger T_\mathbf{k} = 1$. In the diagonal basis the velocity becomes:

$$\tilde{v}_{\mathbf{k},\alpha} \equiv T_\mathbf{k}^\dagger v_{\mathbf{k},\alpha} T_\mathbf{k} = \partial_\alpha\mathcal{E}_\mathbf{k} - i\mathcal{A}_{\mathbf{k},\alpha}\mathcal{E}_\mathbf{k} + i\mathcal{E}_\mathbf{k}\mathcal{A}_{\mathbf{k},\alpha}, \tag{25}$$

where $\mathcal{A}_{\mathbf{k}\alpha} = iT_\mathbf{k}^\dagger\partial_\alpha T_\mathbf{k}$, and $v_{\mathbf{k},\alpha}$ is a Fourier transform of the velocity operator $v_\alpha$.

Let us calculate the $S_{\alpha\beta}$. We denote energy of band $n$ as $(\mathcal{E}_k)_{nn} = \varepsilon_{n\mathbf{k}}$. After some calculations the expression for $S_{\alpha\beta}$ becomes:

$$S_{\alpha\beta} = \frac{i}{2}\sum_{\mathbf{k}n}(\tilde{v}_{\mathbf{k},\alpha})_{nm}\left(\mathcal{E}_\mathbf{k}\tilde{v}_{\mathbf{k},\beta} + \tilde{v}_{\mathbf{k},\beta}\mathcal{E}_\mathbf{k}\right)_{mn}\frac{g\left(\varepsilon_{n\mathbf{k}}\right) - g\left(\varepsilon_{m\mathbf{k}}\right)}{\left(\varepsilon_{n\mathbf{k}} - \varepsilon_{m\mathbf{k}} + i0^+\right)^2} \tag{26}$$

$$= -\frac{i}{2}\left[\sum_{\mathbf{k}n}(\mathcal{A}_{\mathbf{k},\alpha})_{nm}(\mathcal{A}_{\mathbf{k},\beta})_{mn}\varepsilon_{n\mathbf{k}}g\left(\varepsilon_{n\mathbf{k}}\right) + \sum_{\mathbf{k}n}(\mathcal{A}_{\mathbf{k},\alpha})_{nm}\varepsilon_{m\mathbf{k}}(\mathcal{A}_{\mathbf{k},\beta})_{mn}g\left(\varepsilon_{n\mathbf{k}}\right)\right] - (\alpha \leftrightarrow \beta) \tag{27}$$

$$+ \sum_{\mathbf{k}n}(\partial_\alpha\varepsilon_{n\mathbf{k}})(\partial_\beta\varepsilon_{n\mathbf{k}})\varepsilon_{n\mathbf{k}}\frac{1}{2\Gamma_{n\mathbf{k}}}g'(\varepsilon_{n\mathbf{k}}) \tag{28}$$

$$= \frac{i}{2}\sum_{\mathbf{k}n}\int_{-\infty}^\infty d\eta\delta\left[\eta - \varepsilon_{n\mathbf{k}}\right]g(\eta)\left[\left(\partial_\alpha T_\mathbf{k}^\dagger\right)(\eta + H_\mathbf{k})(\partial_\beta T_\mathbf{k})\right]_{nn} - (\alpha \leftrightarrow \beta) + \sum_{\mathbf{k}n}(\partial_\alpha\varepsilon_{n\mathbf{k}})(\partial_\beta\varepsilon_{n\mathbf{k}})\varepsilon_{n\mathbf{k}}\frac{1}{2\Gamma_{n\mathbf{k}}}g'(\varepsilon_{n\mathbf{k}}). \tag{29}$$

We define two distinct by nature terms:

$$S_{\alpha\beta}^{[1]} = \frac{i}{2}\sum_{\mathbf{k}n}\int_{-\infty}^\infty d\eta\delta\left(\eta - \varepsilon_{n\mathbf{k}}\right)g(\eta)\left[\left(\partial_\alpha T_\mathbf{k}^\dagger\right)(\eta + H_\mathbf{k})(\partial_\beta T_\mathbf{k})\right]_{nn} - (\alpha \leftrightarrow \beta), \tag{30}$$

and

$$S_{\alpha\beta}^{[2]} = \sum_{\mathbf{k}n}(\partial_\alpha\varepsilon_{n\mathbf{k}})(\partial_\beta\varepsilon_{n\mathbf{k}})\varepsilon_{n\mathbf{k}}\frac{1}{2\Gamma_{n\mathbf{k}}}g'(\varepsilon_{n\mathbf{k}}). \tag{31}$$

In deriving the $S_{\alpha\beta}^{[2]}$ term we assumed a lifetime of the bosons $\Gamma_{n\mathbf{k}}$, and used the transformations from the previous section for Eq. (13).

The perturbed current is given by the following expression:

$$J_\alpha^{[1]} = \frac{1}{2}\mathrm{Tr}\sum_{\mathbf{k}} (x_\beta v_{\mathbf{k},\alpha} + v_{\mathbf{k},\alpha} x_\beta) g(\mathcal{E}_{\mathbf{k}}) \nabla_\beta \chi. \tag{32}$$

Let us now study the $M_{\alpha\beta}\nabla_\beta\chi$ coefficient, it is defined by the expression:

$$M_{\alpha\beta} = \frac{1}{2}\mathrm{Tr}\sum_{\mathbf{k}} (x_\beta v_{\mathbf{k},\alpha} + v_{\mathbf{k},\alpha} x_\beta) g(\mathcal{E}_{\mathbf{k}}) = \frac{1}{2}\sum_{\mathbf{k}} \int g(\eta)\mathrm{Tr}\left[(x_\beta v_{\mathbf{k},\alpha} - x_\alpha v_{\mathbf{k},\beta})\delta(\eta - H_{\mathbf{k}})\right]d\eta. \tag{33}$$

We follow Smrcka and Streda approach and introduce two functions:

$$A_{\alpha\beta}(\eta) = i\mathrm{Tr}\left[v_{\mathbf{k},\alpha}\frac{dG^{\mathrm{R}}}{d\eta}v_{\mathbf{k},\beta}\delta(\eta - H_{\mathbf{k}}) - v_{\mathbf{k},\alpha}\delta(\eta - H_{\mathbf{k}})v_{\mathbf{k},\beta}\frac{dG^{\mathrm{A}}}{d\eta}\right], \tag{34}$$

$$B_{\alpha\beta}(\eta) = i\mathrm{Tr}\left[v_{\mathbf{k},\alpha}G^{\mathrm{R}}v_{\mathbf{k},\beta}\delta(\eta - H_{\mathbf{k}}) - v_{\mathbf{k},\alpha}\delta(\eta - H_{\mathbf{k}})v_{\mathbf{k},\beta}G^{\mathrm{A}}\right], \tag{35}$$

where $G^{\mathrm{R/A}} = (\eta - H_k \pm i0^+)$. Expression

$$A_{\alpha\beta}(\eta) - \frac{1}{2}\frac{dB_{\alpha\beta}(\eta)}{d\eta} = \frac{1}{4\pi i}\mathrm{Tr}\left[x_\alpha\left(G^{\mathrm{R}}\right)^2 v_{\mathbf{k},\beta} - x_\alpha\left(G^{\mathrm{A}}\right)^2 v_{\mathbf{k},\beta}\right] \tag{36}$$

$$+ \frac{1}{4\pi}\mathrm{Tr}\left[x_\alpha\left(G^{\mathrm{A}} - G^{\mathrm{R}}\right)x_\beta - x_\alpha x_\beta\left(G^{\mathrm{A}} - G^{\mathrm{R}}\right)\right] - (\alpha \leftrightarrow \beta) \tag{37}$$

is useful for further calculations. We also use the following expressions:

$$\left(G^{\mathrm{R}}\right)^2 - \left(G^{\mathrm{A}}\right)^2 = 2\pi i\frac{d}{d\eta}\delta(\eta - H_{\mathbf{k}}), \tag{38}$$

$$\mathrm{Tr}\left\{x_\alpha\left[\left(G^{\mathrm{R}}\right)^2 - \left(G^{\mathrm{A}}\right)^2\right]v_{\mathbf{k},\beta}\right\} - (\alpha \leftrightarrow \beta) = 2\pi i\mathrm{Tr}\left[(x_\alpha v_{\mathbf{k},\beta} - x_\beta v_{\mathbf{k},\alpha})\frac{d}{d\eta}\delta(\eta - H_{\mathbf{k}})\right]. \tag{39}$$

We finally obtain:

$$A_{\alpha\beta}(\eta) - \frac{1}{2}\frac{dB_{\alpha\beta}(\eta)}{d\eta} = \frac{1}{4\pi}\mathrm{Tr}\left[x_\alpha\left(G^{\mathrm{A}} - G^{\mathrm{R}}\right)x_\beta - x_\alpha x_\beta\left(G^{\mathrm{A}} - G^{\mathrm{R}}\right)\right] - (\alpha \leftrightarrow \beta) \tag{40}$$

$$+ \frac{1}{2}\mathrm{Tr}\left[(x_\alpha v_{\mathbf{k},\beta} - x_\beta v_{\mathbf{k},\alpha})\frac{d}{d\eta}\delta(\eta - H_{\mathbf{k}})\right], \tag{41}$$

where the first term after integration over $\eta$ will result in a commutator $[x_\alpha, x_\beta]$, hence will vanish. Vanishing of this commutation relation will only happen when the whole Hilbert space is considered. The following identity for bounded spectrum is of use

$$\int_{-\infty}^{\infty} d\eta \left(A_{\alpha\beta}(\eta) - \frac{1}{2}\frac{dB_{\alpha\beta}(\eta)}{d\eta}\right) = i\int_{-\infty}^{\infty} d\eta \mathrm{Tr}\left[v_{\mathbf{k},\alpha}\frac{dG^{\mathrm{R}}}{d\eta}v_{\mathbf{k},\beta}\delta(\eta - H_{\mathbf{k}}) - v_{\mathbf{k},\alpha}\delta(\eta - H_{\mathbf{k}})v_{\mathbf{k},\beta}\frac{dG^{\mathrm{A}}}{d\eta}\right] \tag{42}$$

$$= -i\int_{-\infty}^{\infty} d\eta \sum_n \delta[\eta - \varepsilon_{n\mathbf{k}}]\left\{\tilde{v}_{\mathbf{k},\alpha}\frac{1}{[\varepsilon_{n\mathbf{k}} - \mathcal{E}_{\mathbf{k}}]^2}\tilde{v}_{\mathbf{k},\beta}\right\}_{nn} - (\alpha \leftrightarrow \beta) \tag{43}$$

$$= -i\sum_n \int_{-\infty}^{\infty} d\eta \delta[\eta - (H_{\mathbf{k}})_{nn}](\mathcal{A}_{\mathbf{k},\beta})_{nm}(\mathcal{A}_{\mathbf{k},\alpha})_{mn} - (\alpha \leftrightarrow \beta) \tag{44}$$

$$= i\sum_n \left[\left(\partial_\beta T_{\mathbf{k}}^\dagger\right)(\partial_\alpha T_{\mathbf{k}})\right]_{nn} - (\alpha \leftrightarrow \beta) = 0. \tag{45}$$

In the expression above $\Omega_{\alpha\beta}^{(n)}(\mathbf{k}) \equiv i\left[\left(\partial_\alpha T_{\mathbf{k}}^\dagger\right)(\partial_\beta T_{\mathbf{k}})\right]_{nn} - (\alpha \leftrightarrow \beta)$ is the $k$-space Berry curvature of the $n$th band. We observe a sum rule, $\sum_n \Omega_{\alpha\beta}^{(n)}(\mathbf{k}) = 0$, which will be uesful in further derivations. Another identity

$$B_{\alpha\beta}(\eta) = i\sum_n \left[\left(\partial_\alpha T_{\mathbf{k}}^\dagger\right)(\eta - H_{\mathbf{k}})(\partial_\beta T_{\mathbf{k}})\right]_{nn}\delta[\eta - \varepsilon_{n\mathbf{k}}] - (\alpha \leftrightarrow \beta) \tag{46}$$

is of use in further derivations. After all these transformations, one can show that

$$M_{\alpha\beta} = \sum_{\mathbf{k}} \left( \int_0^\infty d\eta \int_\eta^\infty d\tilde{\eta} + \int_{-\infty}^0 d\eta \int_\eta^{-\infty} d\tilde{\eta} \right) g(\eta) \left( A_{\alpha\beta}(\tilde{\eta}) - \frac{1}{2} \frac{dB_{\alpha\beta}(\tilde{\eta})}{d\tilde{\eta}} \right) \tag{47}$$

$$= \sum_{\mathbf{k}} \int_{-\infty}^\infty d\tilde{\eta} \left( A_{\alpha\beta}(\tilde{\eta}) - \frac{1}{2} \frac{dB_{\alpha\beta}(\tilde{\eta})}{d\tilde{\eta}} \right) \int_0^{\tilde{\eta}} d\eta g(\eta). \tag{48}$$

We then get a Berry curvature contribution to the response written as

$$S_{\alpha\beta}^{[1]} + M_{\alpha\beta} = -i \sum_{\mathbf{k}n} \int_{-\infty}^\infty d\tilde{\eta} \delta\left[\tilde{\eta} - \varepsilon_{n\mathbf{k}}\right] \left[ \left(\partial_\alpha T_{\mathbf{k}}^\dagger\right) \left(\partial_\beta T_{\mathbf{k}}\right) \right]_{nn} \left[ \int_0^{\tilde{\eta}} g(\eta) d\eta - \tilde{\eta} g(\tilde{\eta}) \right] - (\alpha \leftrightarrow \beta) \tag{49}$$

$$= T \sum_{\mathbf{k}n} \Omega_{\alpha\beta}^{(n)}(\mathbf{k}) c_1(\varepsilon_{n\mathbf{k}}), \tag{50}$$

where $c_1(x) = \int_0^x \eta \frac{dg(\eta)}{d\eta} d\eta - \int_0^\infty \eta \frac{dg(\eta)}{d\eta} d\eta$, in which a sum rule derived above in expression (42), namely $\sum_n \Omega_{\alpha\beta}^{(n)}(\mathbf{k}) = 0$, can be used to eliminate the constant terms of the integral. Finally, the full expression for the response is given by two compact terms

$$J_\alpha = \frac{1}{V} \left[ k_{\rm B} T \sum_{\mathbf{k}n} \Omega_{\alpha\beta}^{(n)}(\mathbf{k}) c_1(\varepsilon_{n\mathbf{k}}) + \sum_{\mathbf{k}n} (\partial_\alpha \varepsilon_{n\mathbf{k}})(\partial_\beta \varepsilon_{n\mathbf{k}}) \varepsilon_{n\mathbf{k}} \frac{1}{2\Gamma_{n\mathbf{k}}} g'(\varepsilon_{n\mathbf{k}}) \right] \nabla_\beta \chi, \tag{51}$$

where we restored $k_{\rm B}$ factor and volume.

### C. Torque response to temperature gradient

In order to calculate the torque response to the temperature gradient, we define two contributions:

$$\langle \partial_{\mathbf{m}} H \rangle \equiv \langle \partial_{\mathbf{m}} H \rangle_{\rm ne} + \frac{1}{2} \langle \partial_{\mathbf{m}} \left[ r_\beta H + H r_\beta \right] \rangle_{\rm eq} \nabla_\beta \chi. \tag{52}$$

The first term is described by the Kubo formula, $\langle \partial_{\mathbf{m}} H \rangle_{\rm ne} = S_{\mathbf{m}\beta} \nabla_\beta \chi$, the second term is defined as $M_{\mathbf{m}\beta} = \frac{1}{2} \langle \partial_{\mathbf{m}} \left[ r_\beta H + H r_\beta \right] \rangle_{\rm eq}$, we then formally rewrite the expression for torque as

$$\langle \partial_{\mathbf{m}} H \rangle = (S_{\mathbf{m}\beta} + M_{\mathbf{m}\beta}) \nabla_\beta \chi = L_{\mathbf{m}\beta} \nabla_\beta \chi. \tag{53}$$

Calculations for the torque are similar to the ones presented for the particle current. The final result is given by the expression:

$$\langle \partial_{\mathbf{m}} H \rangle = \frac{1}{V} \left[ k_{\rm B} T \sum_{\mathbf{k}n} \Omega_{\mathbf{m}\beta}^{(n)}(\mathbf{k}) c_1(\varepsilon_{n\mathbf{k}}) + \sum_{\mathbf{k}n} (\partial_{\mathbf{m}} \varepsilon_{n\mathbf{k}})(\partial_\beta \varepsilon_{n\mathbf{k}}) \varepsilon_{n\mathbf{k}} \frac{1}{2\Gamma_{n\mathbf{k}}} g'(\varepsilon_{n\mathbf{k}}) \right] \nabla_\beta \chi, \tag{54}$$

where now $\Omega_{\mathbf{m}\beta}^{(n)}(\mathbf{k}) \equiv i \left[ \left(\partial_{\mathbf{m}} T_{\mathbf{k}}^\dagger\right) \left(\partial_\beta T_{\mathbf{k}}\right) \right]_{nn} - (\mathbf{m} \leftrightarrow \beta)$ is the mixed space Berry curvature of the $n$th band, and we restored the $k_{\rm B}$ and volume factors.

---



\end{document}